\documentclass[twocolumn]{aastex61}
\usepackage{amsmath}
\usepackage{ulem}

\shorttitle{$A$--$X$ (0,0) bands of C$_2$ and CN}
\shortauthors{Hamano et al.}

\begin{document}

\title{First detection of $A$--$X$ (0,0) bands of interstellar C$_2$ and CN}

\correspondingauthor{Satoshi Hamano}

\author{Satoshi Hamano}
\affiliation{National Astronomical Observatory of Japan, 2-21-1 Osawa, Mitaka, Tokyo 181-8588, Japan}
\affiliation{Laboratory of Infrared High-resolution Spectroscopy (LiH), Koyama Astronomical Observatory, Kyoto Sangyo University, Motoyama, Kamigamo, Kita-ku, Kyoto 603-8555, Japan}
\email{satoshi.hamano@nao.ac.jp}
\author{Hideyo Kawakita}
\affiliation{Laboratory of Infrared High-resolution Spectroscopy (LiH), Koyama Astronomical Observatory, Kyoto Sangyo University, Motoyama, Kamigamo, Kita-ku, Kyoto 603-8555, Japan}
\affiliation{Department of Astrophysics and Atmospheric Sciences, Faculty of Sciences, Kyoto Sangyo University, Motoyama, Kamigamo, Kita-ku, Kyoto 603-8555, Japan}
\author{Naoto Kobayashi}
\affiliation{Kiso Observatory, Institute of Astronomy, School of Science,The University of Tokyo, 10762-30 Mitake, Kisomachi, Kisogun, Nagano, 397-0101, Japan}
\affiliation{Institute of Astronomy, School of Science, University of Tokyo, 2-21-1 Osawa, Mitaka, Tokyo 181-0015, Japan}
\affiliation{Laboratory of Infrared High-resolution Spectroscopy (LiH), Koyama Astronomical Observatory, Kyoto Sangyo University, Motoyama, Kamigamo, Kita-ku, Kyoto 603-8555, Japan}
\author{Keiichi Takenaka}
\affiliation{Laboratory of Infrared High-resolution Spectroscopy (LiH), Koyama Astronomical Observatory, Kyoto Sangyo University, Motoyama, Kamigamo, Kita-ku, Kyoto 603-8555, Japan}
\affiliation{Department of Astrophysics and Atmospheric Sciences, Faculty of Sciences, Kyoto Sangyo University, Motoyama, Kamigamo, Kita-ku, Kyoto 603-8555, Japan}
\author{Yuji Ikeda}
\affiliation{Laboratory of Infrared High-resolution Spectroscopy (LiH), Koyama Astronomical Observatory, Kyoto Sangyo University, Motoyama, Kamigamo, Kita-ku, Kyoto 603-8555, Japan}
\affiliation{Photocoding, 460-102 Iwakura-Nakamachi, Sakyo-ku, Kyoto, 606-0025, Japan}
\author{Noriyuki Matsunaga}
\affiliation{Department of Astronomy, Graduate School of Science, University of Tokyo, Bunkyo-ku, Tokyo 113-0033, Japan}
\affiliation{Laboratory of Infrared High-resolution Spectroscopy (LiH), Koyama Astronomical Observatory, Kyoto Sangyo University, Motoyama, Kamigamo, Kita-ku, Kyoto 603-8555, Japan}
\author{Sohei Kondo}
\affiliation{Kiso Observatory, Institute of Astronomy, School of Science,The University of Tokyo, 10762-30 Mitake, Kiso-machi, Kiso-gun, Nagano, 397-0101, Japan}
\affiliation{Laboratory of Infrared High-resolution Spectroscopy (LiH), Koyama Astronomical Observatory, Kyoto Sangyo University, Motoyama, Kamigamo, Kita-ku, Kyoto 603-8555, Japan}
\author{Hiroaki Sameshima}
\affiliation{Institute of Astronomy, School of Science, University of Tokyo,2-21-1 Osawa, Mitaka, Tokyo 181-0015, Japan}
\affiliation{Laboratory of Infrared High-resolution Spectroscopy (LiH), Koyama Astronomical Observatory, Kyoto Sangyo University, Motoyama, Kamigamo, Kita-ku, Kyoto 603-8555, Japan}
\author{Kei Fukue}
\affiliation{Laboratory of Infrared High-resolution Spectroscopy (LiH), Koyama Astronomical Observatory, Kyoto Sangyo University, Motoyama, Kamigamo, Kita-ku, Kyoto 603-8555, Japan}
\author{Chikako Yasui}
\affiliation{National Astronomical Observatory of Japan, 2-21-1 Osawa, Mitaka, Tokyo 181-8588}
\affiliation{Laboratory of Infrared High-resolution Spectroscopy (LiH), Koyama Astronomical Observatory, Kyoto Sangyo University, Motoyama, Kamigamo, Kita-ku, Kyoto 603-8555, Japan}
\author{Misaki Mizumoto}
\affiliation{Centre for Extragalactic Astronomy, Department of Physics, University of Durham, South Road, Durham DH1 3LE, UK}
\author{Shogo Otsubo}
\affiliation{Laboratory of Infrared High-resolution Spectroscopy (LiH), Koyama Astronomical Observatory, Kyoto Sangyo University, Motoyama, Kamigamo, Kita-ku, Kyoto 603-8555, Japan}
\affiliation{Department of Astrophysics and Atmospheric Sciences, Faculty of Sciences, Kyoto Sangyo University, Motoyama, Kamigamo, Kita-ku, Kyoto 603-8555, Japan}
\author{Ayaka Watase}
\affiliation{Laboratory of Infrared High-resolution Spectroscopy (LiH), Koyama Astronomical Observatory, Kyoto Sangyo University, Motoyama, Kamigamo, Kita-ku, Kyoto 603-8555, Japan}
\affiliation{Department of Astrophysics and Atmospheric Sciences, Faculty of Sciences, Kyoto Sangyo University, Motoyama, Kamigamo, Kita-ku, Kyoto 603-8555, Japan}
\author{Tomohiro Yoshikawa}
\affiliation{Edechs, 17-203 Iwakura-Minamiosagi-cho, Sakyo-ku, Kyoto 606-0003, Japan}
\affiliation{Laboratory of Infrared High-resolution Spectroscopy (LiH), Koyama Astronomical Observatory, Kyoto Sangyo University, Motoyama, Kamigamo, Kita-ku, Kyoto 603-8555, Japan}
\author{Hitomi Kobayashi}
\affiliation{Estrista, 9-205 Iwakura-Minamiosagi-cho, Sakyo-ku, Kyoto 606-0003, Japan}

\begin{abstract}

We report the first detection of C$_2$ $A^1\Pi_u$--$X^1\Sigma_g^+$ (0,0) and CN $A^2\Pi_u$--$X^2\Sigma^+$ (0,0) absorption bands in the interstellar medium. The detection was made using the near-infrared (0.91--1.35 $\mu$m) high-resolution ($R=20,000$ and 68,000) spectra of Cygnus OB2 No.\,12 collected with the WINERED spectrograph mounted on the 1.3 m Araki telescope. The $A$--$X$ (1,0) bands of C$_2$ and CN were detected simultaneously. These near-infrared bands have larger oscillator strengths, compared with the $A$--$X$ (2,0) bands of C$_2$ and CN in the optical. In the spectrum of the C$_2$ (0,0) band with $R=68,000$, three velocity components in the line of sight could be resolved and the lines were detected up to high rotational levels ($J''\sim20$). By analyzing the rotational distribution of C$_2$, we could estimate the kinetic temperature and gas density of the clouds with high accuracy. Furthermore, we marginally detected weak lines of $^{12}$C$^{13}$C for the first time in the interstellar medium. Assuming that the rotational distribution and the oscillator strengths of the relevant transitions of $^{12}$C$_2$ and $^{12}$C$^{13}$C are the same, the carbon isotope ratio was estimated to be $^{12}\text{C}/^{13}\text{C}=50$--100, which is consistent with the ratio in the local interstellar medium. We also calculated the oscillator strength ratio of the C$_2$ (0,0) and (1,0) bands from the observed band strengths. Unfortunately, our result could not discern theoretical and experimental results because of the uncertainties. High-resolution data to resolve the velocity components will be necessary for both bands in order to put stronger constraints on the oscillator strength ratios.




\end{abstract}



\section{Introduction}

C$_2$ and CN are important not only for understanding the chemical evolution of the translucent clouds but also for probing for the physical properties of interstellar clouds \citep{sno06}. Because pure rotational electric dipole transitions of C$_2$ are forbidden because of the lack of permanent electric dipole moments, C$_2$ can be rotationally excited to higher levels by the collisions with atoms and molecules, as well as through electronic transitions by the interstellar radiation field. Therefore, the rotational distribution of C$_2$ can be used for estimating the kinetic temperature and gas density of interstellar clouds \citep{van82,cas12}. On the other hand, a heteronuclear diatomic molecule of CN has a permanent electric dipole moment (i.e., pure rotational transitions are allowed). The rotational excitation temperature of CN has sometimes been used to estimate the brightness temperature of the cosmic microwave background \citep[CMB][]{mey85}. Recently, it was suggested that the slight excess of the rotational excitation temperature from the CMB temperature can mainly be attributed to electron collision and thus can be used as a tracer of electron density \citep{rit11}.

The first detection of interstellar C$_2$ molecules was reported by \citet{sou77}, who detected the (1,0) band of the C$_2$ Phillips system ($A^1 \Pi _u$--$X^1 \Sigma _g^+$) in the line of sight of Cyg OB2 No.\,12. C$_2$ molecules have mainly been investigated with the optical absorption lines of the (2,0) and (3,0) bands at around 8765 and 7720 \r{A}, respectively. The (1,0) and (0,0) Phillips bands located in the near-infrared (NIR) region are not used despite having larger oscillator strengths than the (2,0) and (3,0) bands. This is most likely because it has been difficult to resolve the bands and to detect weak lines at high rotational levels without a high-efficiency and high-resolution NIR spectrograph that has only recently been developed. Moreover, the C$_2$ (0,0) Phillips band is contaminated by many telluric absorption lines, which make it difficult to detect each individual rotational line. In particular, the C$_2$ (0,0) Phillips band has never been detected in the interstellar medium. The situation for CN is similar to that of C$_2$. CN has (1,0) and (0,0) bands of the red system ($A^2 \Pi _u$--$X^2 \Sigma ^+$) in the NIR region. Although these bands are stronger than the (2,0) red band of CN around 7900 \r{A} in the optical region, the lines of the (1,0) and (0,0) bands are contaminated by telluric absorption lines as in the case of the C$_2$ (0,0) Phillips band. The CN (0,0) red band has also never been detected in the interstellar medium.

In this paper, we report the first detection of the C$_2$ (0,0) Phillips band and the CN (0,0) red band in the interstellar medium. These bands were detected in the high-resolution NIR spectra (0.91--1.35 $\mu$m) of Cyg OB2 No.\,12, which is a representative object for the study of the interstellar medium due to its high visual extinction ($A_V \sim 10$ mag) and extreme luminosity \citep{whi15}. The spectra were obtained by the WINERED spectrograph mounted on the 1.3 m Araki telescope. The (1,0) bands of C$_2$ and CN were also detected with high accuracy in our spectra. 
Also, we marginally detected the absorption lines of the (0,0) band of $^{12}$C$^{13}$C. The detection of $^{12}$C$^{13}$C in the interstellar medium has never been reported. 
The rest of this paper is organized as follows. Section 2 describes our observations and data reduction process. Section 3 gives a description of the detected C$_2$ and CN bands. Section 4 discusses the rotational distribution of the C$_2$, the constraints on the oscillator strengths of the C$_2$ bands from our observation, and the carbon isotope ratio. Section 5 gives a summary of this paper. The wavelength of standard air is used throughout this paper.


\section{Observation and Data reduction}

Data were collected with the high-resolution NIR echelle spectrograph, WINERED \citep{ike16}, mounted on the F$/$10 Nasmyth focus of the 1.3 m Araki telescope at Koyama Astronomical Observatory, Kyoto Sangyo University, Japan \citep{yos12}. WINERED uses a 1.7 $\mu$m cutoff $2048\times2048$ HAWAII-2RG infrared array with a pixel scale of 0$''$.8 pixel$^{-1}$. 
WINERED provides three observational modes, WIDE, HIRES-Y, and HIRES-J. The WIDE mode covers 0.91--1.35 $\mu$m range with a spectral resolving power of $R\equiv \lambda/\Delta \lambda =$ 28,000 or $\Delta v =11$ km s$^{-1}$. The HIRES-Y and HIRES-J modes cover the whole $Y$ and $J$ bands, respectively, with a spectral resolving power of $R=68,000$ \citep{ots16}. We already reported the WIDE-mode spectrum of Cyg OB2 No.\,12 in \citet{ham16}, in which diffuse interstellar bands (DIBs) were investigated.

We obtained the NIR spectra of Cyg OB2 No.\,12 using WIDE and HIRES-J modes. Table \ref{targets} shows the observational information, as well as the bands of C$_2$ and CN covered in the observing modes. All of the data were obtained through dithering the telescope by 30$''$ (so-called ABBA sequence). For the removal of telluric absorption lines, the telluric standard A-type stars were also observed at airmasses similar to the target airmasses (see Table \ref{targets}). 

The collected data were reduced with our pipeline software (S. Hamano et al. 2019 in preparation). Using this pipeline, the obtained raw images were reduced to one-dimensional spectra. Once the spectra were obtained, we divided the spectra of Cyg OB2 No.\,12 with the spectra of the corresponding telluric standard stars. Here, we used the IRAF\footnote{IRAF is distributed by the National Optical Astronomy Observatories, which are operated by the Association of Universities for Research in Astronomy, Inc., under cooperative agreement with the National Science Foundation.}/telluric task, which can adjust the strength of the telluric absorption lines based on Beer's law with the single layer atmosphere model and the wavelength shift of the standard star's spectrum. See \citet{sam18} for a full description of our telluric absorption correction method. Then, we fitted a Legendre polynomial function to the continuum regions around the CN and C$_2$ bands by masking absorption lines of these bands and the region where the telluric transmittance was lower than 0.7. The 5th and 10th order Legendre polynomials were used for fitting continuum regions around the CN and C$_2$ bands, respectively. 

Comparing the WIDE-mode spectra obtained at A and B positions on the slit, broad and shallow spurious features are found to be present in 10130--10150 \r{A} only in the spectra obtained at the A position. The features overlap on a part of the C$_2$ (1,0) Phillips band. Therefore, we use only the spectra obtained at the B position for the narrow wavelength region of 10130--10150 \r{A}.  Because the number of available frames is halved, the signal-to-noise ratio (S/N) becomes lower by a factor of about $\sqrt{2}$ for the wavelength region.

\begin{deluxetable*}{ccccccccccc}
\tabletypesize{\scriptsize}
\tablecaption{Observation summary of Cyg OB2 No.\,12 \label{targets}}
\tablehead{
 \colhead{Obs. Date} & \colhead{$R$}  & \multicolumn{3}{c}{Telluric \tablenotemark{a}} & \colhead{} & \multicolumn{2}{c}{Int. Time (s)} &\colhead{S/N\tablenotemark{b}} & C$_2$\tablenotemark{c} & CN\tablenotemark{c}  \\ \cline{3-5} \cline{7-8}
  \colhead{(UT)} & \colhead{}  & \colhead{Name} & \colhead{Sp. Type} & \colhead{$J$ (mag)} & \colhead{} & \colhead{Object} & \colhead{Telluric} & \colhead{Object/Telluric} & \colhead{} & \colhead{}
}
\startdata
 2014 Oct 17 & 20,000 & HR 196  & A2V & 5.291 & & 3600  & 8400 & 700 & (1,0), (0,0) & (1,0), (0,0) \\
 2016 Jul 23 & 68,000 & 29 Vul  & A0V & 5.153 & & 2400  & 3600 & 450 & (0,0) & \nodata  \\
 2016 Aug 1 & 68,000 & HR 8358  & A0V & 5.607 & & 2400  & 7200 & 420 & (0,0) & \nodata    \\
\enddata
\tablecomments{}
\tablenotetext{a}{Telluric standard stars used to correct the telluric absorption lines.}
\tablenotetext{b}{The average signal-to-noise ratio per pixel of the spectrum after the division by the telluric standard spectrum.}
\tablenotetext{c}{The bands included in the wavelength coverage of each observation.}
\end{deluxetable*}

\section{Results}
\subsection{C$_2$ Phillips bands}

\begin{figure*}
\includegraphics[width=18cm,clip]{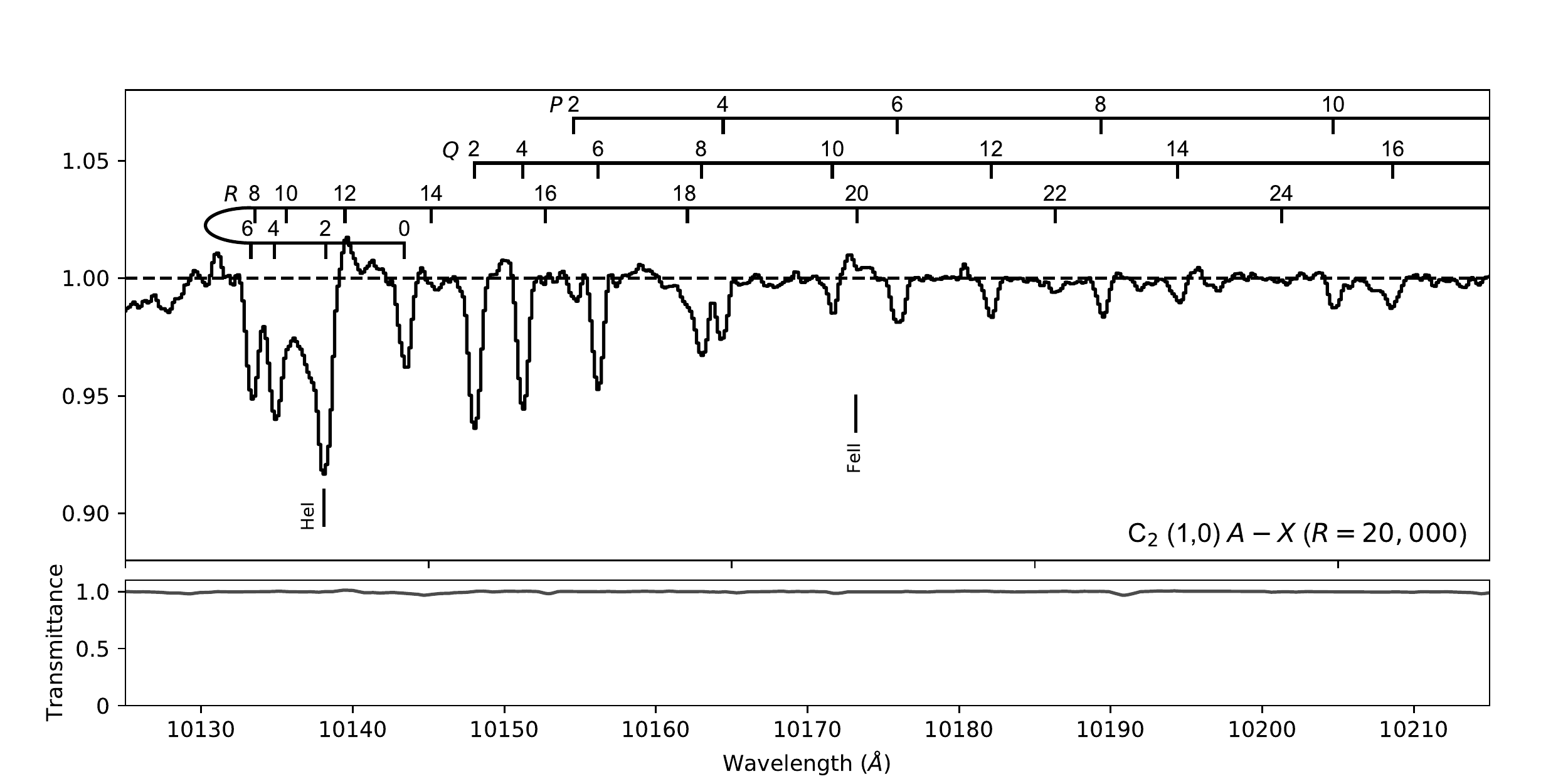}
\includegraphics[width=18cm,clip]{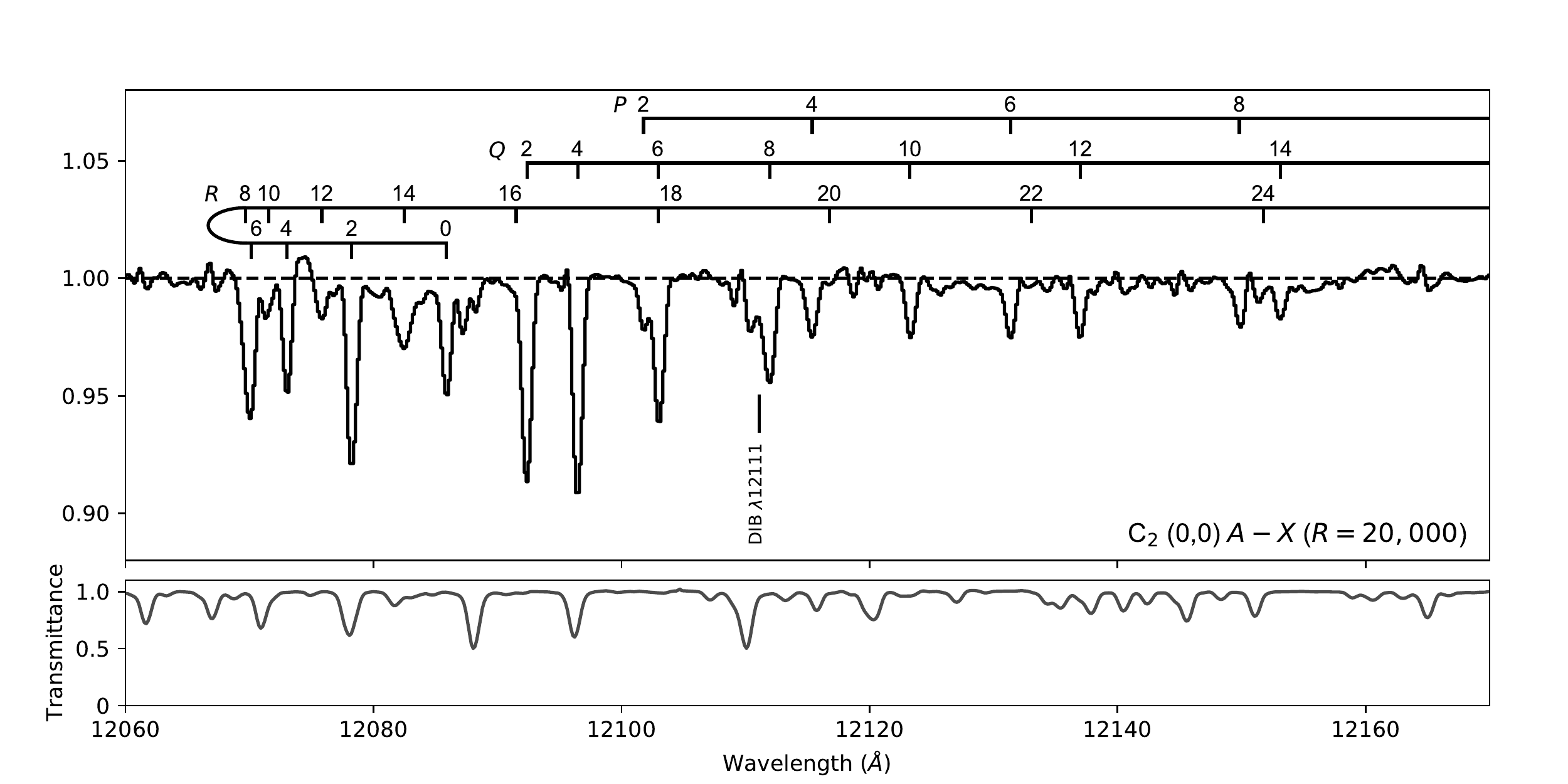}
\caption{Normalized spectra of the C$_2$ (1,0) (upper panel) and (0,0) (lower panel) Phillips bands toward Cyg OB2 No\,12 obtained with the WIDE mode ($R=20,000$). The normalized spectra of the telluric standard stars are also shown. The wavelengths of each line of the bands are marked with lines above the spectrum. The positions of stellar lines and DIBs (including its candidates) are also shown with lines below the spectrum. The wavelengths of the (1,0) and (0,0) bands were calculated from the wavenumbers listed in table 2 in \citet{cha77} and table 1 in \citet{dou88}, respectively.}
\label{wide_c2}
\end{figure*}

\begin{figure*}
\includegraphics[width=18cm,clip]{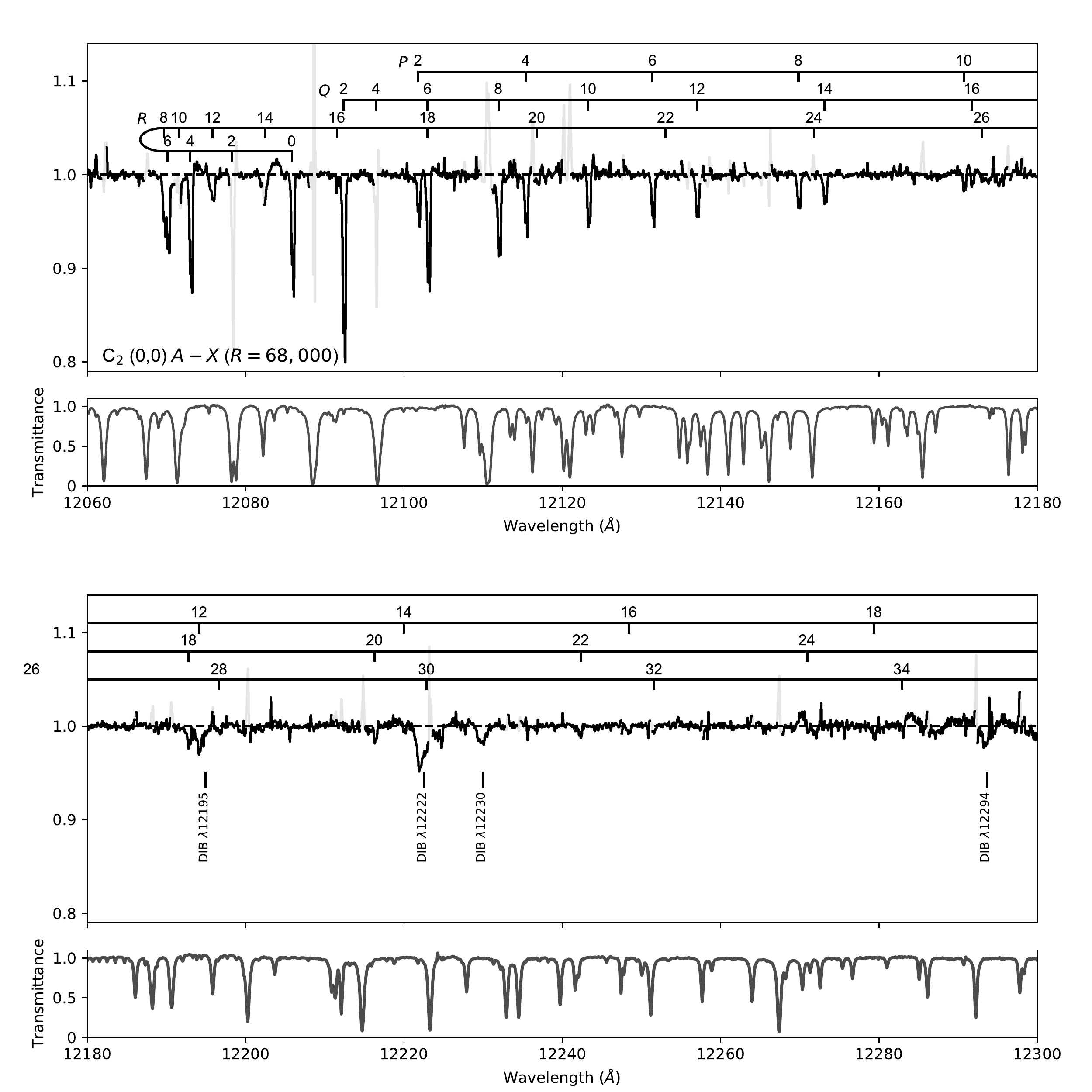}
\caption{Normalized spectrum of the C$_2$ (0,0) Phillips band toward Cyg OB2 No.\,12 obtained with the HIRES-J mode ($R=68,000$). The wavelength ranges at which the transmittance was lower than 0.5 are plotted with gray lines. The stellar lines and DIBs are also marked with lines below the spectrum. The normalized spectrum of the telluric standard star is also shown.}
\label{hires-j_c2}
\end{figure*}

Figures \ref{wide_c2} and \ref{hires-j_c2} show the spectra of C$_2$ bands obtained with WIDE and HIRES-J modes, respectively. Both the C$_2$ (0,0) and (1,0) Phillips bands were clearly detected up to the rotational levels of $J'' > 20$ in the ground state for both the WIDE and HIRES-J spectra. The stellar absorption lines and DIBs, with which the C$_2$ bands are contaminated, are marked in the figures. These DIBs are newly found in this observation and will be reported in another paper (S. Hamano et al. 2019 in preparation). The telluric lines are very weak in the (1,0) band region while the (0,0) band is contaminated with strong telluric absorption lines. The absorption lines of telluric water vapor in the HIRES-J spectrum were particularly strong because the HIRES-J data were obtained in the summer season when the temperature and humidity at the observational site are high. Therefore, as for the HIRES-J data, the wavelength ranges at which the atmospheric transmittance was lower than 0.5 (plotted with gray lines in Figure \ref{hires-j_c2}) were not used in the following analysis of the C$_2$ band.

\begin{figure}
\includegraphics[width=8cm,clip]{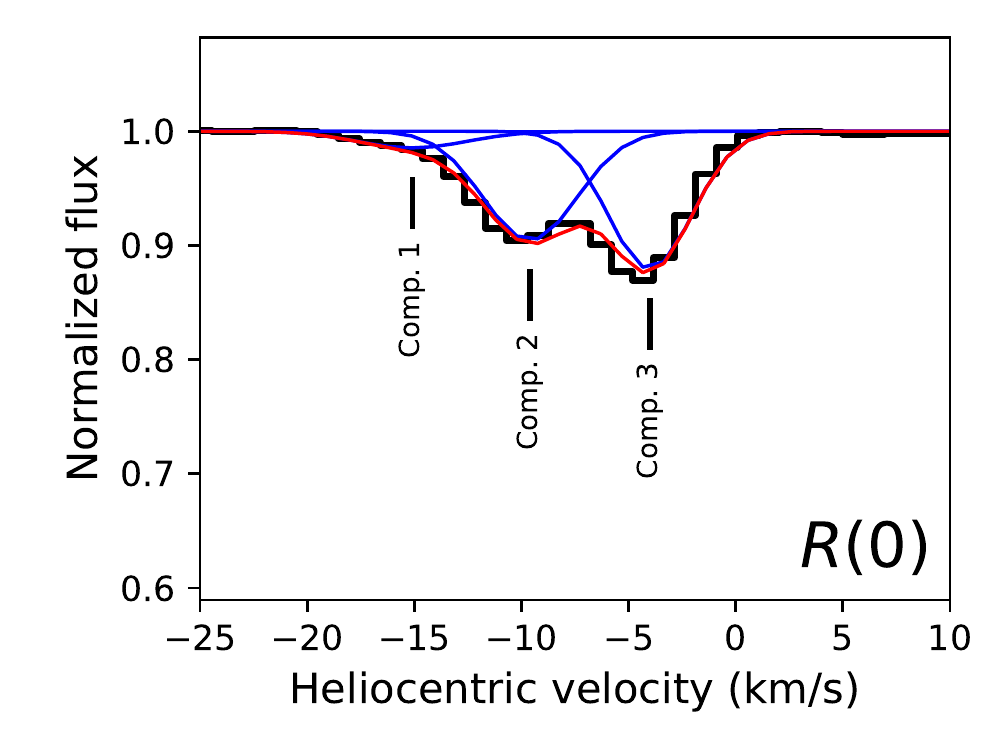}
\includegraphics[width=8cm,clip]{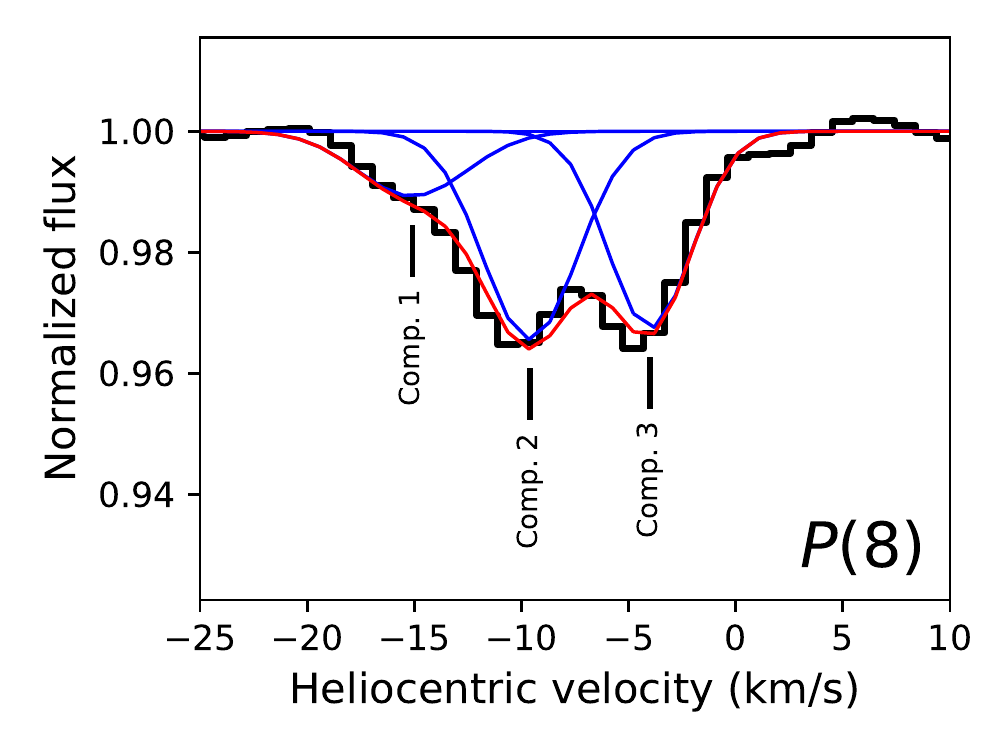}
\caption{Close-up images of the spectrum of the $R(0)$ and $P(8)$ lines in the C$_2$ (0,0) band. The black lines show the spectrum of Cyg OB2 No.\,12 obtained with the HIRES-J mode. Three components at $-15.1$, $-9.6$, and $-4.0$ km s$^{-1}$ were detected. The blue lines show the Gaussian curves fitted to the spectrum. The red lines show the composite of the Gaussian curves.}
\label{hires-j_c2_zoom}
\end{figure}

Figure \ref{hires-j_c2_zoom} shows the close-up plots for the $R(0)$ and $P(8)$ lines of the C$_2$ (0,0) band, which were clearly detected in the HIRES-J spectrum. Three velocity components were resolved with $R=68,000$. In this paper, the three components are referred to as components 1, 2, and 3 as shown in Figure \ref{hires-j_c2_zoom}. These three components in the line of sight of Cyg OB2 No.\,12 were already recognized by \citet{mcc02} with the CO and \ion{K}{1} absorption spectra. They also detected the C$_2$ (2,0) Phillips band, but could not detect the absorption lines at the velocity of component 1. Components 1, 2, and 3 could not be resolved in our WIDE spectrum ($\Delta v = 15$ km s$^{-1}$). 

We estimated the line-of-sight velocities, Doppler widths, and column densities of each velocity component using VoigtFit \citep{kro18}, which is a Python package for fitting Voigt profiles to absorption lines. Because the line parameters of C$_2$ Phillips bands are not included in VoigtFit, we originally calculated the oscillator strengths of each rotational line of C$_2$ bands. 
We used the oscillator strengths of $f_{00}=2.233 \times 10^{-3}$ and $f_{10}=2.348 \times 10^{-3}$ calculated by \citet{sch07}. The oscillator strengths of each line were calculated from the following relation \citep{gre01}: 
\begin{equation}
f_{J' J''} = f_{\text{band}} \frac{\nu _{J' J''}}{\nu_ {\text{band}}} \frac{S_{J' J''}}{2(2J''+1)},
\end{equation}
where $f_{\text{band}}$ is the oscillator strength of the band ($f_{00}$ or $f_{10}$), $\nu_ {\text{band}}$ is the wavenumber of the vibrational band, and  $S_{J' J''}$ are H\"onl-London factors: $(J''+2)$, $(2J'' + 1)$, and $(J''-1)$ for the $R$, $Q$, and $P$ branches, respectively. 
The wavelengths of the C$_2$ (0,0) and (1,0) bands were adopted from \citet{dou88} and \citet{cha77}, respectively.

The C$_2$ lines were fitted simultaneously to the HIRES-J spectrum using VoigtFit. The regions contaminated with strong telluric absorption lines were eliminated for the fitting as well as for the estimate of the S/N. The S/N of each pixel was calculated considering the loss of flux by the moderate telluric absorption lines. The C$_2$ lines blended with other features, such as stellar lines and DIBs, were not included in the fitting. 

Table \ref{ew_hires-j} lists the column densities for each velocity component obtained by the fitting. The heliocentric velocities of components 1,2 and 3 were estimated as $-15.1$, $-9.6$, and $-4.0$ km s$^{-1}$, respectively. The Doppler widths of each velocity component could be estimated in the fitting because the EWs of the strongest $P$, $Q$, and $R$ lines in the \deleted{saturated} optically thick regime. The Doppler widths of components 1, 2, and 3 were estimated as $2.39 \pm 0.42$, $1.47 \pm 0.15$ and $0.53 \pm 0.06$ km s$^{-1}$, respectively. We did not conduct the Voigt profile analysis for the WIDE data because the three velocity components, could not be resolved with $R=20,000$ ($\Delta v = 15$ km s$^{-1}$). Therefore, we used $N(J'')$ determined from the HIRES-J data in the following analysis.

According to \citet{mcc02}, who obtained the optical spectrum of Cyg OB2 No.\,12 with $R=200,000$, the FWHMs of the C$_2$ (2,0) band absorption lines were measured to be $3.0\pm0.2$ and $2.0\pm0.2$ km s$^{-1}$ for components 2 and 3, respectively. Considering the broadening effect by instrumental profiles with $\Delta v = 1.5$ km s$^{-1}$, the Doppler widths estimated from our fitting, $1.47 \pm 0.15$ and $0.53 \pm 0.06$ km s$^{-1}$ for component 2 and 3, correspond to FWHMs of 2.9$\pm0.3$ and $1.7\pm0.1$ km s$^{-1}$, respectively, in the $R=200,000$ spectrum. These values were consistent with the directly measured FWHMs by \citet{mcc02}.

The column densities obtained by our analysis were compared with those of \citet{gre01} and \citet{mcc02}, both of which analyzed the spectrum of the (2,0) Phillips band of Cyg OB2 No.\,12. \citet{gre01} obtained the spectrum with $R=45,000$ and S/N $>$ 600 and could not resolve the velocity components while \citet{mcc02} resolved the velocity components with $R=200,000$ but the S/N\footnote{\citet{mcc02} did not show the value of S/N for their C$_2$ spectrum of Cyg OB2 No.\,12. Judging from the spectrum shown in their paper, the S/N appears to be about 100.} was not as good as that of \citet{gre01}. Although both of them assumed the optically thin conditions for the calculation of column densities considering the typical Doppler width of $b=1$ km s$^{-1}$ \citep{gre01}, this assumption is not valid in the case of $b=0.5$ km s$^{-1}$. Therefore, we calculated the column densities from the (2,0) band equivalent widths (EWs) by \citet{gre01} and \citet{mcc02} using the Doppler widths and the velocities estimated from our fitting, and then compared the resultant column densities with those we derived from the (0,0) band. We adopted $f_{20} = 1.424 \times 10^{-3}$ \citep{sch07} for the calculation. The ratios of the column densities from the (2,0) band to our values from the (0,0) band were calculated up to $J'' = 8$, below which we could detect the absorption lines of the velocity component 1. The ratios of the column density of \citet{gre01} to our value was $0.94 \pm 0.15$. The column densities calculated from the data of \citet{gre01} were close to our results but the uncertainties were large. Note that the standard deviation does not include the uncertainties of EWs and Doppler widths. As for \citet{mcc02}, the ratios were calculated to be $0.79\pm 0.07$ and $0.74 \pm 0.07$ for components 2 and 3, respectively, meaning that the EWs of \citet{mcc02} were lower than those expected from the (0,0) band. Considering the relatively low S/N of the \citet{mcc02} spectrum, the discrepancy could be attributed to the systematic noise, which could not be estimated in their paper \citep[see the footnote of Table 6 in][]{mcc02}. Also, the systematic uncertainties due to the continuum normalization could be large because the strong lines in the (2,0) band are detected within the broad stellar lines of \ion{H}{1} Paschen 12 and \ion{He}{1}. Therefore, the column densities obtained from our (0,0) band data could be more robust than the previous results because of both the high quality of our data and intrinsically large EWs of the (0,0) band. \deleted{Therefore, the column densities obtained from the (0,0) band could be more robust than the previous results. }

To measure the EWs of the detected lines independently, we fitted Gaussian curves to the spectrum data. The velocities and widths were fixed in the fit as determined by the VoigtFit analysis for the HIRES-J spectrum. For the partially blended lines, multiple Gaussian curves at the wavelengths of the blended lines were fitted simultaneously. While the EWs of the three velocity components were measured separately in the HIRES-J spectrum, the total EWs of the three components were measured in the WIDE spectrum. For the EW uncertainties, we included the systematic uncertainties from continuum fitting estimated with the rms shift method \citep{sem92} in addition to the statistical uncertainties. Tables \ref{ew_hires-j} and \ref{ew_wide} list the EWs measured for the absorption lines identified as C$_2$ Phillips bands in the spectra obtained with the HIRES-J and WIDE modes, respectively. 
In the WIDE spectrum, the lower spectral resolution caused the blending of some multiple C$_2$ lines and it made it difficult to normalize the spectrum using the surrounding continuum regions. We did not list the EWs for those lines, whose systematic uncertainties of EWs were much larger due to the blending.
In Table \ref{ew_wide}, we also list the EWs of the (1,0) band calculated from $N(J'')$ and $b$ determined by HIRES-J data using $f_{10} = 2.348 \times 10^{-3}$ \citep{sch07}. Comparing the measured and calculated EWs, we discuss the ratio of observationally constrained oscillator strengths of Phillips bands in Section 4.2. Note that the total EWs measured from WIDE and HIRES spectra are consistent within uncertainties for the most lines detected in both spectra.

\begin{deluxetable*}{ccccccccccccc}
\tabletypesize{\scriptsize}
\tablecaption{Equivalent widths and column densities for the C$_2$ $(0,0)$ Phillips Bands in the HIRES-J mode spectrum of Cyg OB2 No.\,12 \label{ew_hires-j}}
\tablehead{
 \colhead{} & \colhead{}  & \colhead{} & \colhead{} & \multicolumn{2}{c}{Comp 1}   & \colhead{}  & \multicolumn{2}{c}{Comp 2}   & \colhead{}  &\multicolumn{2}{c}{Comp 3} & \colhead{}   \\ \cline{5-6} \cline{8-9} \cline{11-12}
 \colhead{$J''$} & \colhead{Branch}  & \colhead{$\lambda_\text{air}$ (\r{A})} & \colhead{$f (10^{-3})$}  & \colhead{EW (m\r{A})} & \colhead{$N$ ($10^{12}$ cm$^{-2}$) \tablenotemark{a}} & \colhead{}  & \colhead{EW (m\r{A})} & \colhead{$N$ ($10^{12}$ cm$^{-2}$) \tablenotemark{a}} & \colhead{}  & \colhead{EW (m\r{A})}  & \colhead{$N$ ($10^{12}$ cm$^{-2}$) \tablenotemark{a}}  & \colhead{Blending\tablenotemark{b}} 
}
\startdata
$0$	& $R$ & 12086.244 & 2.230 & $3.55 \pm 0.40$ & $0.93 \pm 0.21$ & & $20.55 \pm 0.31$ & $7.40 \pm 0.24$ & & $23.93 \pm 0.33$ & $10.83 \pm 0.33$ & \\
$2$	& $R$ & 12078.631 & 0.894 & \nodata & $2.94 \pm 0.55$ & & \nodata & $27.25 \pm 0.60$ & & \nodata & $39.63 \pm 1.66$ & Telluric \\
		& $Q$ & 12092.725 & 1.120  & $5.51 \pm 0.41$ &  & & $36.13 \pm 0.32$ &  & & $36.82 \pm 0.34$ &  & \\
		& $P$ & 12102.141 & 0.223  & $2.09 \pm 0.40$ &  & & $8.54 \pm 0.31$ &  & & $10.81 \pm 0.33$ &  & \\
$4$	& $R$ & 12073.398 & 0.745 & $6.47 \pm 0.36$ & $5.75 \pm 0.70$ & & $24.16 \pm 0.28$ & $25.69 \pm 0.82$ & & $24.01 \pm 0.29$ & $30.96 \pm 0.74$ & \\
		& $Q$ & 12096.879 & 1.120  & \nodata &  & & \nodata &  & & \nodata &  & Telluric \\
		& $P$ & 12115.742 & 0.371  & $2.64 \pm 0.41$ &  & & $10.81 \pm 0.33$ &  & & $12.56 \pm 0.35$ &  & \\
$6$	& $R$ & 12070.540 & 0.688 & \nodata & $3.52 \pm 0.47$ & & $15.72 \pm 0.32$ & $17.90 \pm 0.54$ & & $13.91 \pm 0.30$ & $20.15 \pm 0.47$ & $R(8)$ \\
		& $Q$ & 12103.413 & 1.110  & \nodata &  & & \nodata &  & & \nodata &  & $R(18)$ \\
		& $P$ & 12131.760 & 0.428  & $2.79 \pm 0.33$ &  & & $9.72 \pm 0.26$ &  & & $10.60 \pm 0.28$ &  & \\
$8$	& $R$ & 12070.055 & 0.658 & $2.47 \pm 0.36$ & $3.89 \pm 0.40$ & & $10.07 \pm 0.28$ & $12.46 \pm 0.46$ & & \nodata & $12.34 \pm 0.29$ & $R(6)$ \\
		& $Q$ & 12112.335 & 1.110  & $6.25 \pm 0.40$ &  & & $17.11 \pm 0.31$ &  & & $15.29 \pm 0.33$ &  & \\
		& $P$ & 12150.211 & 0.457  & $3.03 \pm 0.41$ &  & & $7.70 \pm 0.31$ &  & & $6.56 \pm 0.33$ &  & \\
$10$	& $R$ & 12071.943 & 0.639 & \nodata & \nodata & & \nodata & $9.58 \pm 0.40$ & & \nodata & $8.19 \pm 0.34$ & Telluric \\
		& $Q$ & 12123.652 & 1.110  & \nodata &  & & $13.88 \pm 0.27$ &  & & $10.52 \pm 0.28$ &  & \\
		& $P$ & 12171.114 & 0.475  & \nodata &  & & $5.10 \pm 0.22$ &  & & $4.30 \pm 0.23$ &  & \\
$12$	& $R$ & 12076.208 & 0.626 & $5.07 \pm 0.35$ & $2.26 \pm 0.38$ & & $5.61 \pm 0.27$ & $7.27 \pm 0.39$ & & $4.66 \pm 0.29$ & $6.18 \pm 0.32$ & \\
		& $Q$ & 12137.378 & 1.110  & $2.81 \pm 0.34$ &  & & $10.20 \pm 0.27$ &  & & $7.96 \pm 0.29$ &  & \\
		& $P$ & 12194.491 & 0.487  & \nodata &  & & \nodata &  & & \nodata &  & DIB \\
$14$	& $R$ & 12082.853 & 0.616 & \nodata & $1.70 \pm 0.34$ & & \nodata & $4.33 \pm 0.70$ & & \nodata & $3.78 \pm 0.29$ & Telluric \\
		& $Q$ & 12153.526 & 1.110  & $3.14 \pm 0.41$ &  & & $6.52 \pm 0.31$ &  & & $5.35 \pm 0.33$ &  & \\
		& $P$ & 12220.370 & 0.495  & \nodata &  & & $1.98 \pm 0.39$ &  & & $1.72 \pm 0.41$ &  & \\
$16$	& $R$ & 12091.888 & 0.609 & \nodata & \nodata & & $2.48 \pm 0.33$ & $3.25 \pm 0.28$ & & $1.32 \pm 0.33$ & $2.01 \pm 0.25$ & \\
		& $Q$ & 12172.113 & 1.110  & \nodata &  & & $4.73 \pm 0.22$ &  & & $2.95 \pm 0.23$ &  & \\
		& $P$ & 12248.774 & 0.501  & \nodata &  & & $1.82 \pm 0.31$ &  & & $1.24 \pm 0.34$ &  & \\
$18$	& $R$ & 12103.322 & 0.603 & \nodata & $1.24 \pm 0.33$ & & \nodata & $2.78 \pm 0.32$ & & \nodata & $2.07 \pm 0.28$ & $Q(6)$ \\
		& $Q$ & 12193.162 & 1.110  & $1.73 \pm 0.40$ &  & & $4.18 \pm 0.30$ &  & & $2.73 \pm 0.33$ &  & \\
		& $P$ & 12279.735 & 0.505  & \nodata &  & & \nodata &  & & \nodata &  & \\
$20$	& $R$ & 12117.168 & 0.598 & \nodata & \nodata & & $2.16 \pm 0.32$ & $2.72 \pm 0.26$ & & $2.88 \pm 0.33$ & $2.34 \pm 0.23$ & \\
		& $Q$ & 12216.696 & 1.100  & \nodata &  & & $3.59 \pm 0.39$ &  & & $2.08 \pm 0.41$ &  & \\
		& $P$ & 12313.287 & 0.508  & \nodata &  & & $2.31 \pm 0.22$ &  & & $1.55 \pm 0.24$ &  & \\
$22$	& $R$ & 12133.443 & 0.593 & \nodata & \nodata & & \nodata & $1.94\pm 0.34$ & & \nodata & \nodata & \\
		& $Q$ & 12242.736 & 1.100  & \nodata &  & & $2.75 \pm 0.31$ &  & & \nodata &  & \\
		& $P$ & 12349.467 & 0.510  & \nodata &  & & \nodata &  & & \nodata &  & \\
$24$	& $R$ & 12152.165 & 0.589 & \nodata & \nodata & & \nodata & \nodata & & \nodata & \nodata & Telluric \\
		& $Q$ & 12271.315 & 1.100  & \nodata &  & & \nodata &  & & \nodata &  & \\
		& $P$ & 12388.311 & 0.511  & \nodata &  & & \nodata &  & & \nodata &  & \\
$26$	& $R$ & 12173.354 & 0.586 & \nodata & \nodata & & $1.36 \pm 0.22$ & $1.40 \pm 0.30$ & & $0.78 \pm 0.23$ & $1.36 \pm 0.27$ & \\
		& $Q$ & 12302.463 & 1.100  & \nodata &  & & $1.83 \pm 0.22$ &  & & $2.07 \pm 0.24$ &  & \\
		& $P$ & 12429.866 & 0.512  & \nodata &  & & \nodata &  & & \nodata &  & \\
$28$	& $R$ & 12197.033 & 0.582 & \nodata & \nodata & & \nodata & \nodata & & $1.35 \pm 0.33$ & $1.53 \pm 0.27$ & \\
		& $Q$ & 12336.214 & 1.090  & \nodata &  & & \nodata &  & & $2.06 \pm 0.28$ &  & \\
		& $P$ & 12474.185 & 0.512  & \nodata &  & & \nodata &  & & \nodata &  & \\
\enddata
\tablecomments{The symbol "\nodata'' denotes undetected lines.}
\tablenotetext{a}{The column densities of the rotational level $J''$ estimated from the simultaneous profile fit to the $P$, $Q$, and $R$ lines.}
\tablenotetext{b}{The features blended with the lines. "Telluric'' means the blending of the strong telluric lines.}
\end{deluxetable*}

\begin{deluxetable}{ccccc}
\tabletypesize{\scriptsize}
\tablecaption{Equivalent widths for the C$_2$ (0,0) and (1,0) Phillips Bands in the WIDE-mode spectrum of the Cyg OB2 No.\,12 \label{ew_wide}}
\tablehead{
 \colhead{} & \colhead{}  & \colhead{(0,0)} & \colhead{(1,0)} & \colhead{(1,0) calculation\tablenotemark{a}} \\
 \colhead{$J''$} & \colhead{Branch}  & \colhead{EW (m\r{A})} & \colhead{EW (m\r{A})} & \colhead{EW (m\r{A})}
}
\startdata
$0$ & $R$ & \nodata & $35.30 \pm 1.49$ & 35.93 \\
$2$ & $R$ & $66.36 \pm 1.27$ & \nodata & 49.57 \\
         & $Q$ & $75.55 \pm 0.89$ & $60.35 \pm 1.50$ & 59.54 \\
         & $P$ & $19.29 \pm 0.94$ & \nodata & 14.09 \\
$4$ & $R$ & \nodata & \nodata & 39.39 \\
         & $Q$ & \nodata & \nodata & 56.16 \\
         & $P$ & $23.59 \pm 1.12$ & $20.49 \pm 1.02$ & 20.87 \\
$6$ & $R$ & \nodata & \nodata & 25.26 \\
         & $Q$ & \nodata & $41.35 \pm 1.49$ & 39.44 \\
         & $P$ & $19.78 \pm 1.13$ & $17.31 \pm 1.18$ & 16.26 \\
$8$ & $R$ & \nodata & \nodata &17.17 \\
         & $Q$ & \nodata & $27.26 \pm 0.95$ & 28.47 \\
         & $P$ & $17.34 \pm 0.97$ & $12.15 \pm 1.09$ & 12.23 \\
$10$ & $R$ & \nodata & \nodata & 11.14 \\
         & $Q$ & $22.26 \pm 1.13$ & $20.57 \pm 1.17$ & 19.17 \\
         & $P$ & \nodata & \nodata & 8.48 \\
$12$ & $R$ & \nodata & \nodata & 9.16 \\
         & $Q$ & $19.33 \pm 1.13$ & $13.57 \pm 1.09$ & 16.18 \\
         & $P$ & \nodata & $6.33 \pm 0.98$ & 7.29 \\
$14$ & $R$ & \nodata & \nodata & 5.69 \\
         & $Q$ & $11.97 \pm 0.96$ & \nodata & 10.28 \\
         & $P$ & \nodata & \nodata & 4.67 \\
$16$ & $R$ & \nodata & \nodata & 3.45 \\
         & $Q$ & \nodata & \nodata & 6.34 \\
         & $P$ & \nodata & \nodata &2.90 \\
$18$ & $R$ & \nodata & $5.05 \pm 0.97$ &3.47 \\
         & $Q$ & \nodata & \nodata &6.41 \\
         & $P$ & \nodata & \nodata &2.99 \\
$20$ & $R$ & $3.53 \pm 1.12$ & \nodata & 3.18 \\
         & $Q$ & \nodata & \nodata &5.92 \\
         & $P$ & \nodata & \nodata &2.78 \\
\enddata
\tablecomments{}
\tablenotetext{a}{EWs calculated from $N(J'')$ in Table \ref{ew_hires-j}.}
\end{deluxetable}

\subsection{CN red band}

\begin{deluxetable*}{ccccccccc}
\tabletypesize{\scriptsize}
\tablecaption{Equivalent widths and column densities for the CN red bands in the WIDE-mode spectrum of Cyg OB2 No.\,12 \label{cn_summary}}
\tablehead{
 \colhead{Band} & \colhead{Transition} & \colhead{$N''$} & \colhead{$J''$} & \colhead{$\lambda _{\text{air}}$\tablenotemark{a}} & \colhead{$f _{\text{osc}}$\tablenotemark{a,b}} & \colhead{EW} & \colhead{$N(N'')$\tablenotemark{c}} & \colhead{$N(N'')$\tablenotemark{d}} \\
\colhead{} &\colhead{} & \colhead{} & \colhead{} & \colhead{(\r{A})} & \colhead{($\times 10 ^{-3}$)} & \colhead{(m\r{A})} & \colhead{($10^{12}$ cm$^{-2}$)}  & \colhead{($10^{12}$ cm$^{-2}$)} 
}
\startdata
(1,0) band	& $^{S}R_{21}(0)$ 		& $0$ & $1/2$ 	& 9139.699 	& 0.2646 	& $6 \pm 3$	& $31 \pm 15$ & $31 \pm 15$	\\
	& $R_{2}(1)+^{R}Q_{21}(1)$ 	& $1$ & $1/2 + 3/2$ 	& 9142.848 	& 0.4375  	& $12 \pm 3$	& $37 \pm 9$ & $40 \pm 10$ \\
	& $^{R}Q_{21}(0)$ 			& $0$ & $1/2$ 	& 9144.055 	& 0.6409 	& $17 \pm 3$	& $36 \pm 6$ & $38\pm6$ \\
	& $Q_{2}(1)+^{Q}P_{21}(1)$ 	& $1$ & $1/2 + 3/2$	& 9147.219 	& 0.3203  	& \nodata & \nodata & \nodata \\
	& $R_{1}(1)$ 				& $1$ & $3/2$	& 9183.230 	& 0.6408 	& $14\pm 3$ & $45\pm9$	& $46 \pm 9$ \\
	& $R_{1}(0)$ 				& $0$ & $1/2$ 	& 9186.950 	& 1.0127 	& $21\pm 3$	& $28 \pm 4$ & $29\pm4$ \\
	& $^{Q}R_{12}(1)+Q_{1}(1)$ 	& $1$ & $1/2 + 3/2$ 	& 9190.138 	& 0.5851  	& $20\pm 3$	& $46\pm 7$ & $50\pm8$ \\
(0,0) band 	& $^SR_{21}(0)$ 	& $0$ & $1/2$ 	& 10925.147 	& 0.3251 	& $10 \pm 2$	& $29 \pm 6$ & $29 \pm 6$ \\
	& $R_{2}(1)+^{R}Q_{21}(1)$ 	& $1$ & $1/2 + 3/2$ 	& 10929.650 	& 0.5386  	& 	\nodata \tablenotemark{e} & \nodata  & \nodata 	\\
	& $^{R}Q_{21}(0)$ 			& $0$ & $1/2$ 	& 10931.442 	& 0.7892 	& 	\nodata \tablenotemark{e} & \nodata   & \nodata  	\\
	& $^{Q}P_{21}(1)+Q_{2}(1)$ 	& $1$ & $1/2 + 3/2$	& 10935.970 	& 0.3945   & 	\nodata  \tablenotemark{e} & \nodata   & \nodata  	\\
	& $R_{1}(1)$ 				& $1$ & $3/2$ 	& 10987.395 	& 0.7886 	& $17\pm 2$	& $30\pm3$ & $33\pm3$ \\
	& $R_{1}(0)$ 				& $0$ & $1/2$ 	& 10992.869 	& 1.2461 	& $74\pm 2$	& $55\pm 2$ & $67 \pm 3$ \\
	& $Q_{1}(1)+^{Q}R_{12}(1)$ 	& $1$ & $1/2 + 3/2$ 	& 10997.445 	& 0.7198  	& \nodata \tablenotemark{f}	& \nodata  & \nodata \\
\enddata
\tablecomments{}
\tablenotetext{a}{The values were adopted from \citet{bro14}.}
\tablenotetext{b}{The effective oscillator strengths are shown for the blended lines.\deleted{The weighted average of the oscillator strengths are shown for the blended lines.}}
\tablenotetext{c}{The column densities calculated with optically thin assumption.}
\tablenotetext{d}{The column densities calculated with the assumption of the cloud parameters from C$_2$ bands (see the text for details).}
\tablenotetext{e}{Overlapped with the stellar Pa $\gamma$ line.}
\tablenotetext{f}{Overlapped with the stellar \ion{He}{1} line.}
\end{deluxetable*}

\begin{figure*}
\includegraphics[width=9.5cm,clip]{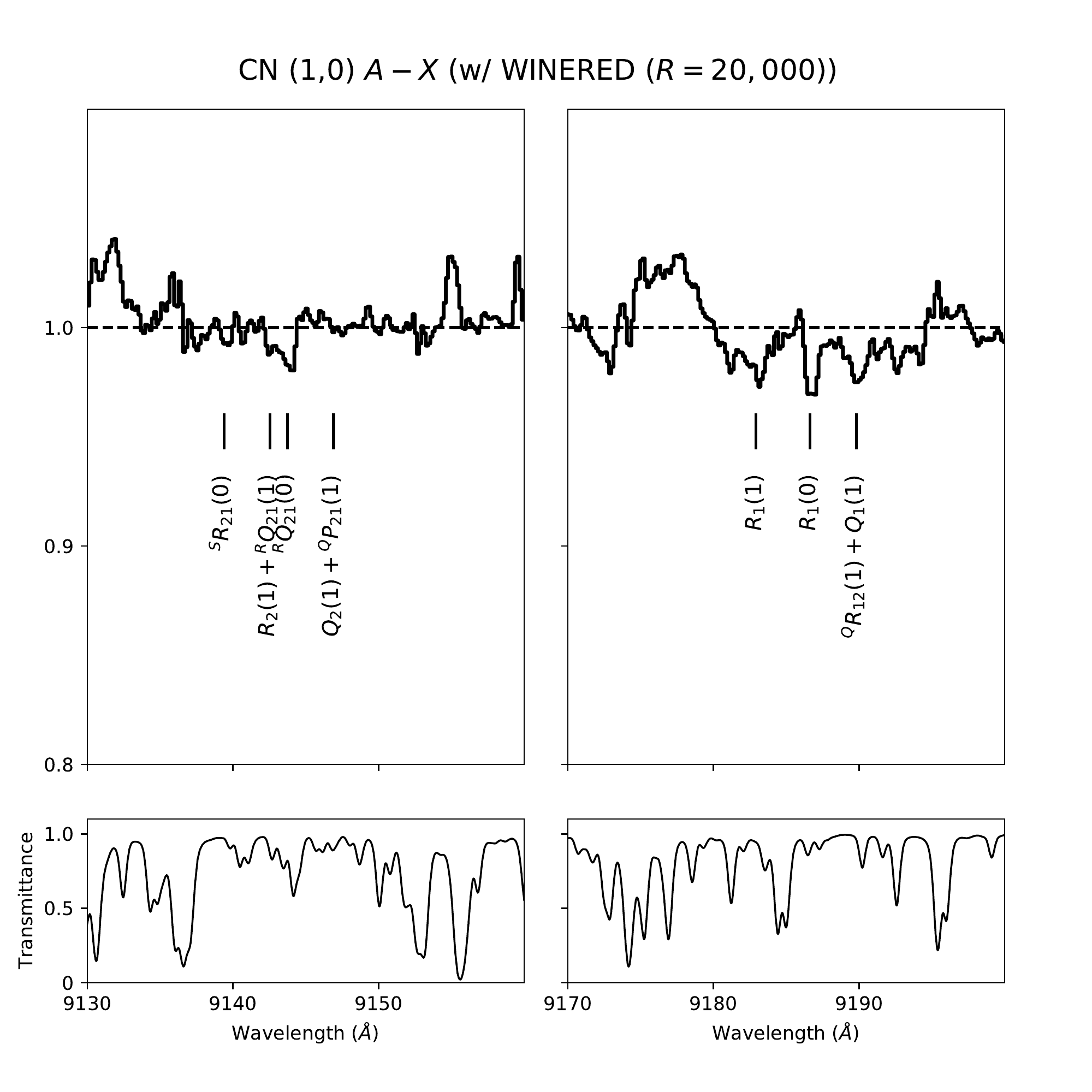}
\includegraphics[width=9.5cm,clip]{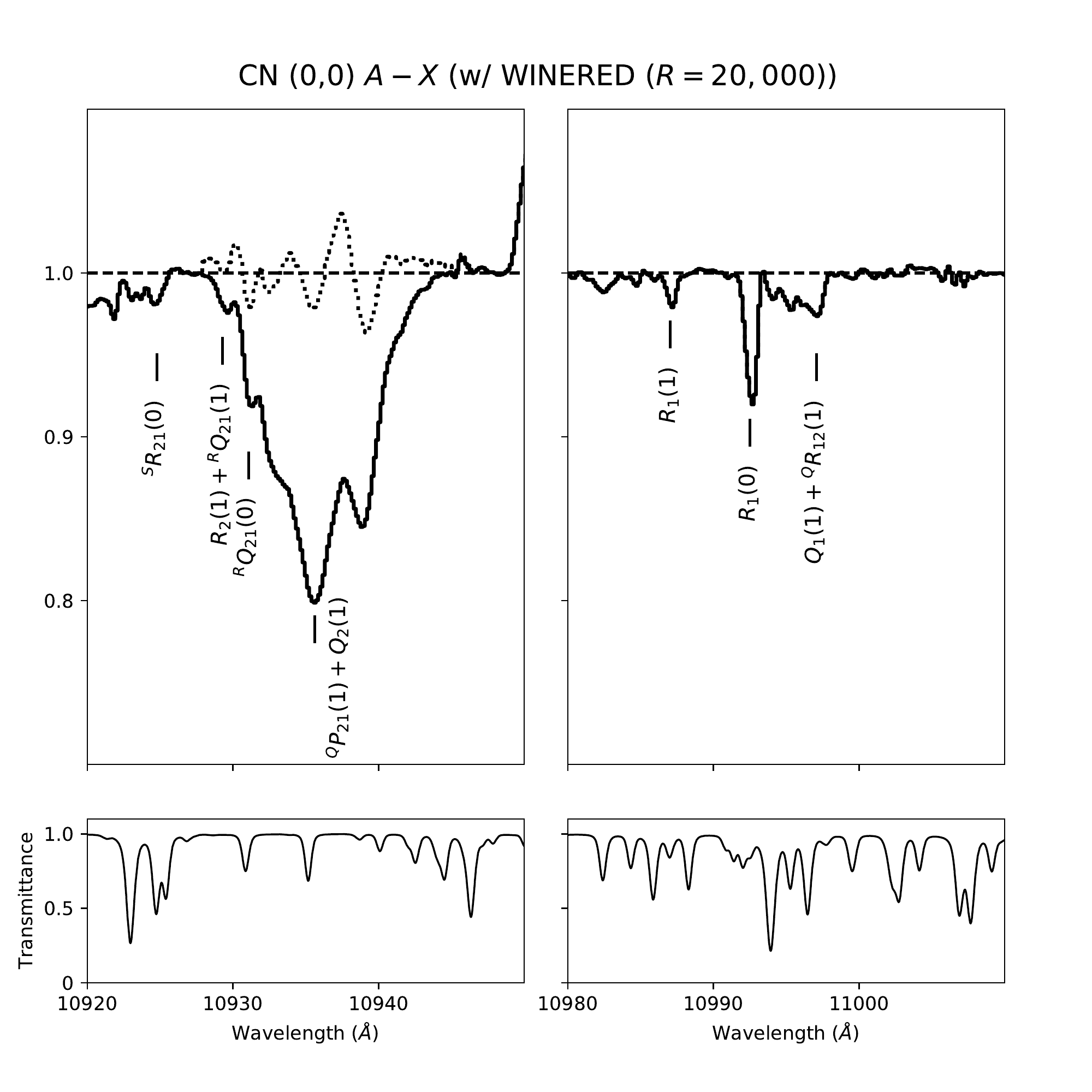}
\caption{The left two panels and right two panels show the spectra of the CN (1,0) and (0,0) red bands toward Cyg OB2 No.\,12, respectively. The transmittance spectrum synthesized using ATRAN \citep{lor92} is also shown, in the lower panels. The wavelengths of the CN red system are adopted from \citet{bro14}. The dotted line shows the spectrum after subtracting the Gaussian function arbitrarily fitted to the Pa $\gamma$ line.}
\label{cn_no12}
\end{figure*}

Figure \ref{cn_no12} shows the WIDE spectra of the CN (1,0) and (0,0) red bands. The rest-frame wavelengths were adopted from \citet{bro14}. As shown in Figure \ref{cn_no12}, some absorption lines were detected at the wavelengths of CN bands. In particular, the three lines of the (0,0) band around $\lambda=$ 10990 \r{A} were clearly detected. As long as we know, this is the first detection of a (0,0) band of the CN red system in the interstellar medium. The lines around 10930\r{A} are overlapped with the broad absorption line of \ion{H}{1} Pa$\gamma$. Within the broad feature of the stellar \ion{H}{1} line, narrow dips can be seen at the wavelengths of the CN band. These dips are most likely CN lines, but it is almost impossible to measure their EWs accurately. 
Table \ref{cn_summary} lists the EWs and column densities measured for the lines of the CN (1,0) and (0,0) red bands in the spectrum of Cyg OB2 No.\,12. 
The EWs were measured by fitting a Gaussian curve to the spectrum. Although these CN lines should originate from two major velocity components, components 2 and 3, as well as C$_2$, they could not be resolved. The measured EWs would be the sum of all velocity components.

Considering the small Doppler widths measured from the C$_2$ (0,0) band, some detected CN lines are probably optically thick. However, since we could not resolve the velocity components, we could not estimate the Doppler widths of each velocity component. Therefore, we calculated CN column densities based on two assumptions. One is the assumption of the optically thin condition and the other is the assumption that the CN lines that originated from components 1, 2 and 3 and have the same Doppler widths as the C$_2$ lines. In the latter assumption, the ratios of column densities of each velocity component to the total column densities were assumed as those of the C$_2$ column densities.
The oscillator strengths calculated by \citet{bro14} were adopted. For the blended lines from the $N''=1$ level, we calculated the effective oscillator strengths by assuming that the populations of the spin-rotational levels $(N'', J'')$ with the same rotational quantum number $N''$ are populated according to their statistical weights
\deleted{the effective oscillator strengths were calculated by averaging the oscillator strengths of blended lines with the statistical weights used as the weighting factors} 
\citep{gre02}. 
Table \ref{cn_summary} lists the column densities calculated under two assumptions. As for the $R_1(0)$ line of the CN (0,0) band, which is the strongest detected line, the column density was much different by the assumptions. 
The column densities of the $N''=0$ and 1 levels estimated from each line, $N(N''=0)$ and $N(N''=1)$, have larger variance than the uncertainties (Table \ref{cn_summary}). This would be because of the systematic uncertainties due to the surrounding stellar lines, which made it difficult to properly evaluate the appropriate continuum level. Also, both bands were severely contaminated by telluric absorption lines.

The (2,0) and (1,0) bands of the CN red system toward Cyg OB2 No.\,12 were also detected by \citet{gre01}. As for the (1,0) band, which was also detected in this study, \citet{gre01} showed that the EWs of $Q_{2}(1)+^{Q}P_{21}(1)$ and $R_{1}(0)$ were $14.3 \pm 5$\r{A} and $25 \pm 5$\r{A}, respectively, and the EW upper limits of $^{S}R_{21}(0)$ and $R_{1}(1)$ were $<28$\r{A} and $<53$\r{A}, respectively. The EWs and upper limits of the (1,0) band measured here were consistent with those reported by \citet{gre01} within uncertainties.

Averaging the column densities calculated with the optically thin condition assumption listed in Table \ref{cn_summary}, the column densities of the $N''=0$ and 1 levels were estimated to be $N(N''=0)= (4.7 \pm 0.2) \times 10^{13}$ and $N(N''=1)=(3.2 \pm 0.3) \times 10^{13}$ cm$^{-2}$. The rotational excitation temperature and the CN total column density were estimated to be $T_{10}=3.7 \pm 0.3$ K and $N(\text{CN})= (8.2 \pm 0.4) \times 10^{13}$ cm$^{-2}$. Note that these results are likely to be affected by the systematic uncertainties due to the residual of telluric absorption lines and the stellar lines.

\deleted{If we used} Using only the $R_1(0)$ and $R_1(1)$ lines of the CN (0,0) band, which were clearly detected without the contamination of stellar lines, we estimated the rotational excitation temperature and the CN total column density to be $T_{10}=3.0 \pm 0.2$ K and $N($CN$)= (1.01 \pm 0.04) \times 10^{14}$ cm$^{-2}$, respectively, in the optically thick case for the $R_{1}(0)$ and $R_{1}(1)$ lines of CN (0,0) band. We suggest that the latter result would be more robust because the rotational temperature is close to the CMB temperature. \citet{gre01} estimated $N(N''=0) = 4.4 \times 10^{13}$ and $N(N''=1) = 5.5 \times 10^{13}$ cm$^{-2}$ (the oscillator strength was adjusted to the values of Brooke et al. 2014 for comparison). Our result should be more robust than that of \citet{gre01} since the $N(N''=1)$ by \citet{gre01} had a large uncertainty and their excitation temperature, $T_{10}=6.2$ K, was much higher than the CMB temperature. However, there is a systematic uncertainty in our results due to the assumption about the Doppler widths and the CN column density ratios among velocity components. It is necessary to resolve the velocity components with higher resolution and detect more lines to obtain these parameters with high accuracy.

\section{Discussion}

\subsection{Rotational excitation of C$_2$}

The gaseous temperature and density of the collision partner ($n(\text{H}) + n(\text{H}_2)$) can be evaluated from the rotational distribution of C$_2$ molecules. These parameters are essential for understanding of the interstellar chemical reactions \citep{gre01}. Using the model of \citet{van82}, we estimated the parameters for the three velocity components of Cyg OB2 No.\,12 from the column densities of each rotational level in the ground state. We use the updated value of the absorption rates calculated by \citet{van84}. We assumed the scaling factor for the incident radiation field to be $I=1$. As the recent analysis of interstellar C$_2$ excitation by \citet{hup12}, we adopted the C$_2$-H$_2$ collisional cross section as $\sigma _0 = 4 \times 10^{-16}$ cm$^{-2}$, which is indicated by recent calculations \citep{lav91,rob92,naj08,naj09}, rather than the frequently used value of $2 \times 10^{-16}$ cm$^{-2}$ \citep{van82}. Although the lines from $J''>20$ were detected in our observation, the original model by \citet{van82} cannot predict the rotational distributions at $J''>20$, because they calculated the radiation excitation matrix and the quadruple transition probabilities up to only $J''=20$. To extrapolate the model to $J''>20$, we estimated the values of the model parameters at $J''>20$ levels by fitting power-law functions of $J''$, which reproduced the values at $J''\leq20$ well.

\begin{figure}
\includegraphics[width=8cm,clip]{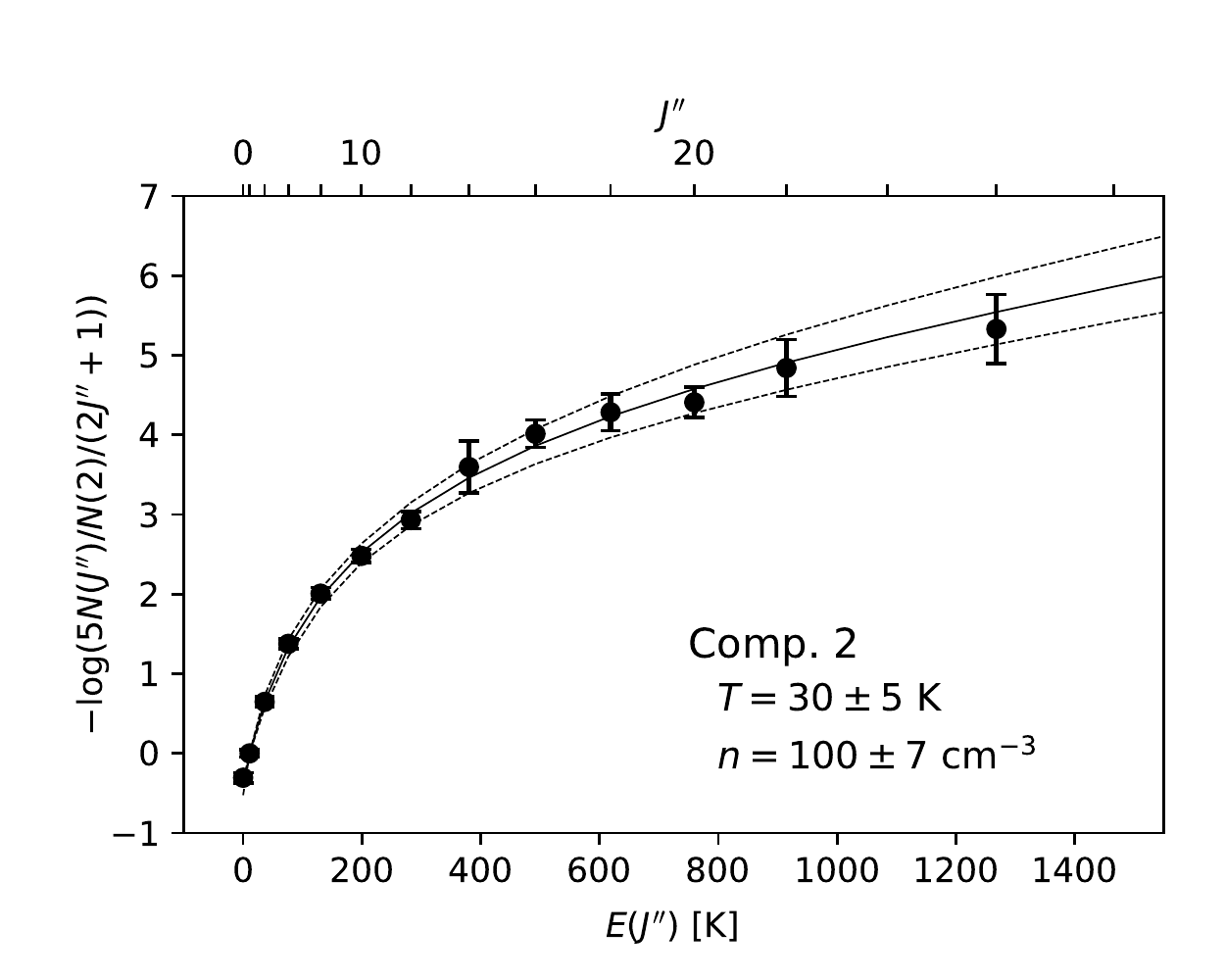}
\includegraphics[width=8cm,clip]{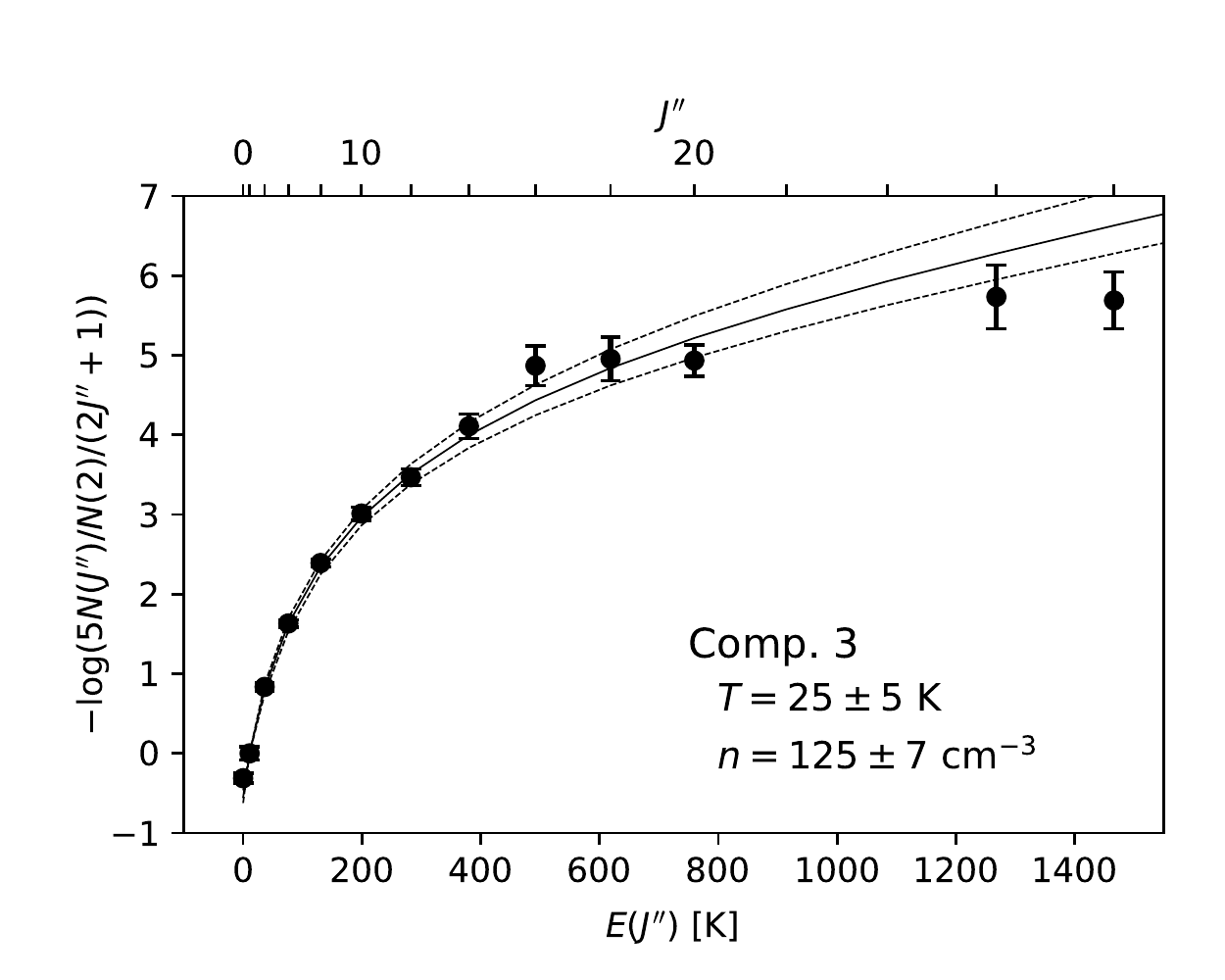}
\caption{The rotational diagrams for components 2 (upper panel) and 3 (lower panel). Best-fit model lines and its one sigma ranges are plotted with the solid and dashed lines, respectively.}
\label{c2_boltzmann}
\end{figure}

Figure \ref{c2_boltzmann} shows the rotational diagrams for each velocity component. The HIRES-J data were used. We could not estimate the parameters for component 1 because of the large uncertainties of column densities. For components 2 and 3, the kinetic temperatures and gas densities $(T,n)$ were estimated as $(30\pm5$ K, $100\pm 7$ cm$^{-3})$ and $(25\pm5$ K, $125\pm 7$ cm$^{-3})$, respectively. We compare these results with the previous results of C$_2$ for Cyg OB2 No.\,12 \citep{gre01,mcc02}. Because both of the previous papers used the collisional cross section of $2 \times 10^{-16}$ cm$^{-2}$, which is the half of our adopted value, their densities were scaled by a factor of about 2 for the comparison in the following. \citet{mcc02} estimated the parameters to be (40 K , 110 cm$^{-3}$) and (30 K, 105 cm$^{-3}$) for components 2 and 3, respectively, but with large uncertainties: $T=25-55$ K and $n = 75 - 300$ cm$^{-3}$ (3$\sigma$ errors). \citet{gre01} analyzed the rotational distribution summed over the components and estimated the parameters to be $(T, n) = (35$ K, $150\pm25$ cm$^{-3}$). They detected the lines of the C$_2$ (2,0) band up to $J''=18$ but could not resolve the velocity components. In comparison with the previous results, both parameters estimated in this paper are consistent with the previous values within uncertainties. Because of the higher oscillator strength of the C$_2$ (0,0) band than the C$_2$ (2,0) band, we could estimate the parameters with an accuracy higher than that of previous studies, even those with similar S/Ns.

From the rotational distribution, the C$_2$ total column densities of components 2 and 3 were estimated to be $(1.29^{+0.16}_{-0.10}) \times 10^{14}$ and $(1.48^{+0.13}_{-0.08}) \times 10^{14}$ cm$^{-2}$, respectively. 
By summing up the column densities listed in Table \ref{ew_hires-j},  the C$_2$ total column densities of components 1, 2, and 3 were $2.2 \times 10^{13}$, $1.2 \times 10^{14}$ and $1.4\times 10^{14}$ cm$^{-2}$, respectively. Table \ref{parameters} summarizes the parameters of the velocity components of Cyg OB2 No.\,12 obtained from the C$_2$ and CN bands.

\begin{deluxetable*}{ccccccccccc}
\tabletypesize{\scriptsize}
\tablecaption{Summary of parameters \label{parameters}}
\tablehead{
\colhead{Comp.} & \colhead{$v$} 
 & \colhead{$b$} & \colhead{$N_\text{obs}$(C$_2$) \tablenotemark{a}} & \colhead{$N_\text{calc}$(C$_2$) \tablenotemark{b}} & \colhead{$T_\text{rot}$(C$_2$)} & \colhead{$n$(C$_2$)} &  \colhead{$N_{J''=0}$(CN)} & \colhead{$N_{J''=1}$(CN)} & \colhead{$T_\text{rot}$(CN)} & \colhead{$N$(CN)} \\
\colhead{} & \colhead{(km s$^{-1}$)} & \colhead{(km s$^{-1}$)} & \colhead{($10^{12}$ cm$^{-2}$)} & \colhead{($10^{12}$ cm$^{-2}$)} & \colhead{(K)} & \colhead{(cm$^{-3}$)} & \colhead{($10^{12}$ cm$^{-2}$)} & \colhead{($10^{12}$ cm$^{-2}$)} & \colhead{(K)} & \colhead{($10^{12}$ cm$^{-2}$)} 
}
\startdata
1 & $-15.1$ & $2.39\pm0.42$ & $22$ & \nodata & \nodata & \nodata & \nodata & \nodata & \nodata & \nodata\\
2 & $-9.6$  & $1.47\pm0.15$ & $120$ & $129^{+16}_{-10}$ & $30\pm5$ & $100\pm 7$ & \nodata & \nodata & \nodata & \nodata\\
3 & $-4.0$  & $0.53\pm0.06$ & $140$ & $148^{+13}_{-8}$ & $25\pm5$ & $125\pm 7$ & \nodata & \nodata & \nodata & \nodata \\
Sum & \nodata & \nodata & 282 & \nodata & \nodata & \nodata & $67 \pm 3$ \tablenotemark{c} & $33 \pm 3$ \tablenotemark{c} & $3.0 \pm 0.2$ & $101 \pm 4$ \\
\enddata
\tablecomments{The values are in units of 10$^{12}$ cm$^{-2}$.}
\tablenotetext{a}{The sum of the column densities of all rotational levels that were measured from detected lines.}
\tablenotetext{b}{The C$_2$ total column densities estimated by fitting the model of \citet{van82} to the observed rotational population.}
\tablenotetext{c}{The column densities estimated from the EWs of $R_1(0)$ and $R_1(1)$ of CN (0,0) band assuming the cloud parameters measured from C$_2$ (0,0) bands (see the text for details).}
\end{deluxetable*}

\subsection{Oscillator strengths}

The oscillator strengths of the C$_2$ Phillips bands have been investigated experimentally and theoretically. However, the experimental and theoretical results are not in agreement with one another. 
Astronomical observations of C$_2$ molecules have put a unique constraint on the oscillator strength ratios of C$_2$ bands \citep{lam95}. Here, we try to put constraints on the oscillator strength ratios of the C$_2$ Phillips bands of Cyg OB2 No.12 using the line strengths of (1,0) and (0,0) bands from this study. We use the (0,0) band in HIRES spectrum and the (1,0) band in WIDE spectrum. It is assumed that the column densities of rotational levels $J''$ in the line of sight of Cyg OB2 No.\,12 are constant between our observations with WIDE and HIRES-J modes. From the strength ratios between the detected lines of (1,0) and (0,0) bands from the same lower level $J''$, the oscillator strength ratio, $f_{00} / f_{10}$, can be constrained. 

Because the strong lines of Cyg OB2 No.\,12 are optically thick for both (0,0) and (1,0) bands (\textsection{3.1}), the oscillator strength ratio cannot be simply calculated from the ratio of the EWs. Therefore, we searched for the oscillator strength ratio that reproduces well the observed EWs with the following procedures. First, we calculated the total EWs of the three velocity components for each rotational lines of the (1,0) band by synthesizing the absorption profiles using the column densities of each rotational level listed in Table \ref{ew_hires-j} and the cloud parameters (the line-of-sight velocities and Doppler widths), which were determined from the (0,0) band (Table \ref{ew_wide}). By varying the oscillator strength of the (1,0) bands from $f_{10} = 2.348 \times 10^{-3}$ \citep{sch07} in the calculation, we searched for the best value of $f_{10}$ reproducing the observed EWs of the (1,0) band. As a result, the $f_{00}/f_{10}$ was estimated to be $0.96 \pm 0.05$. 

The ratios were compared with the experimental results \citep{dav84,bau85} and calculated values \citep{sch07} in Table \ref{osc}. Although the obtained ratio is closer to the theoretical result than the experimental result, it is consistent with both the experimental and theoretical results within the 1$\sigma$ uncertainty. High-resolution data that can resolve the velocity components will be necessary for both bands in order to put stronger constraints on the oscillator strength ratios.
The oscillator strength constraints from astronomical data can contribute to future improvements of the C$_2$ estimate (not only for the oscillator strengths of relevant bands of C$_2$, but also for astrophysical measurements).\deleted{This comparison suggests that the obtained ratios are consistent with the theoretical result rather than the experimental results. This does not mean that our observation supports the absolute value of theoretically-obtained oscillator strengths because we lack information regarding the actual amount of C$_2$ molecules. These oscillator strength constraints would contribute to future improvements of the C$_2$ estimate (not only for the oscillator strengths of relevant bands of C$_2$, but also for astrophysical measurements). High-resolution data to resolve the velocity components will be necessary for both bands in order to put stronger constraints on the oscillator strength ratios.}

\begin{deluxetable*}{ccccc}
\tabletypesize{\scriptsize}
\tablecaption{Oscillator strengths of C$_2$ Phillips bands \label{osc}}
\tablehead{
 \colhead{Reference} & \colhead{Method} & \colhead{$f_{00} \times 10^3$} & \colhead{$f_{10} \times 10^3$} & \colhead{$f_{00} / f_{10}$}
}
\startdata
\citet{dav84} & Experiment & $1.38 \pm 0.14$ & $1.70 \pm 0.17$ & $0.81 \pm 0.11$  \\
\citet{bau85} & Experiment & 1.38 & 1.56 & 0.89 \\
\citet{sch07} & Calculation & 2.233 & 2.348 & 0.951 \\
This study & Astronomy & \nodata & \nodata & $0.96 \pm 0.05$  \\
\enddata
\end{deluxetable*}

\subsection{Marginal detection of $^{12}$C$^{13}$C}

We searched for the lines of the $^{12}$C$^{13}$C (0,0) Phillips band in the HIRES-J spectrum. The wavelengths were based on the wavenumbers measured by \citet{ami83}. $^{12}$C$^{13}$C has the rotational levels of both odd and even rotational quantum numbers $J''$ in the ground state in contrast to $^{12}$C$_2$, which has only even numbers of $J''$ due to the homonuclear molecule nature. As the expected strengths of the $P$ lines were too weak to be detected, only the $Q$ and $R$ lines were searched. The wavelengths of the $^{12}$C$^{13}$C lines are distributed in the range of almost the same wavelength range of $^{12}$C$_2$. Although the $^{12}$C$^{13}$C (1,0) Phillips band and the $^{13}$C$^{14}$N (0,0) and (1,0) red bands covered in the WIDE spectrum were also searched using the line lists from \citet{ami83} and \citet{sne14}, respectively, we could not detect any absorption lines of these bands or put meaningful constraints on the isotope ratio.

\begin{figure}
\includegraphics[width=8.5cm,clip]{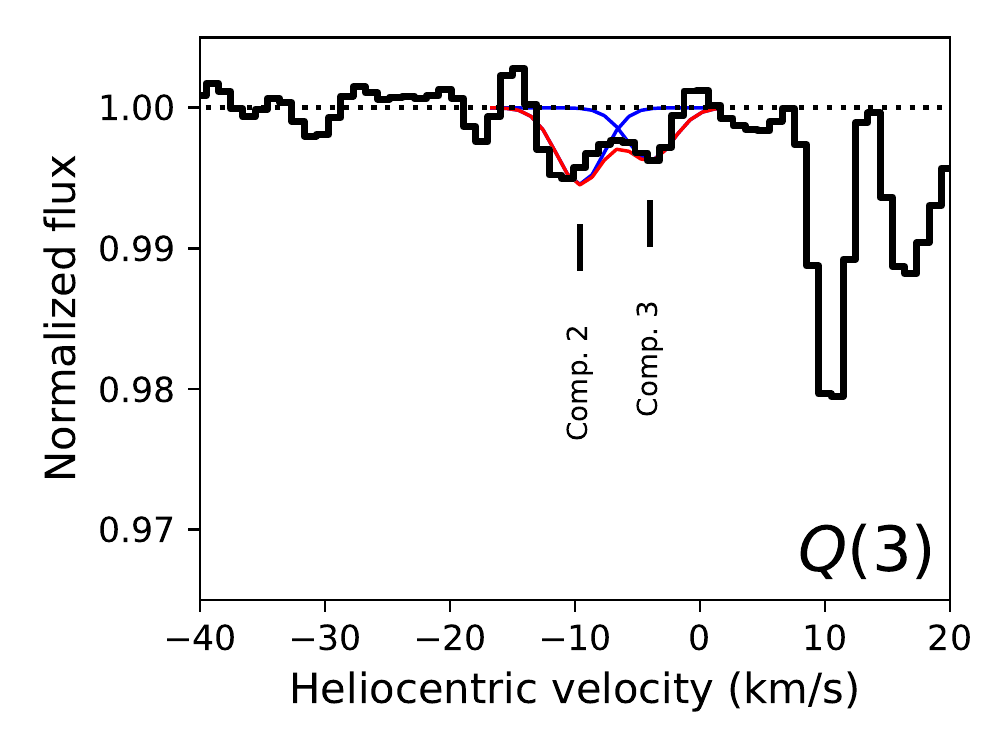}
\caption{Spectrum of the $^{12}$C$^{13}$C $Q(3)$ line. The weak absorption lines were detected at the velocities of components 2 and 3. The absorption lines seen at about +10 and +16 km s$^{-1}$ are the $^{12}$C$_2$ $R(16)$ lines of the components 2 and 3, respectively.}
\label{12C13C_Q3}
\end{figure}

As a result, only $Q(3)$ lines were marginally detected at the velocities of components 2 and 3. To date, the $^{12}$C$^{13}$C molecule has never been detected in the interstellar medium. Figure \ref{12C13C_Q3} shows the spectrum of the $Q(3)$ lines of $^{12}$C$^{13}$C. Their EWs were $1.0\pm0.3$ and $0.7\pm0.3$ m\r{A} for components 2 and 3, respectively. Although these dips were very weak, the velocities matched with those of components 2 and 3 for the $^{12}$C$_2$ lines. Other $Q$ and $R$ lines are expected to be weaker or contaminated with telluric absorption lines and $^{12}$C$_2$ lines; thus, the $Q(3)$ line is the most detectable line in our spectrum of Cyg OB2 No.\,12. The multiple line detection for $^{12}$C$^{13}$C requires higher S/Ns for this target. 

From the EWs of $Q(3)$ lines, we estimated the abundance of $^{12}$C$^{13}$C molecules. We assumed that the oscillator strength and the rotational distribution of $^{12}$C$^{13}$C in the ground state are the same as $^{12}$C$_2$. In contrast to $^{12}$C$_2$, $^{12}$C$^{13}$C can have pure rotational electric dipole transitions, which can cool down the rotational populations of $^{12}$C$^{13}$C, as they have very weak permanent electric dipole moment. We do not have any information regarding the rotational distribution of $^{12}$C$^{13}$C, and thus we ignore the pure rotational electric dipole transition of $^{12}$C$^{13}$C. Because the probability is estimated to be very low \citep{kri87}, the difference of rotational distributions between $^{12}$C$_2$ and $^{12}$C$^{13}$C would not be so large. However, note that \citet{bak98}, who detected the $^{12}$C$^{13}$C absorption band in a circumstellar shell, found that the lower rotational excitation temperature of $^{12}$C$^{13}$C, compared with that of $^{12}$C$_2$, is probably due to the radiative cooling of $^{12}$C$^{13}$C by the pure rotational electric dipole transition. 

On the basis of these assumptions, the $^{12}$C$_2$ / $^{12}$C$^{13}$C ratios for Cyg OB2 No.\,12 were estimated to be $\sim25$ and $\sim50$ for components 2 and 3, respectively. Conversion of these values into carbon isotope ratios results in $^{12}$C / $^{13}$C $\sim50$ and $\sim100$, respectively. Considering the large uncertainties in the EWs, this is roughly consistent with the carbon isotope ratios in the interstellar medium measured by various carbon molecules: $68\pm 15$ at the solar galactocentric distance \citep[measurements of CO, CN, and H$_2$CO in][]{mil05} and $76\pm2$ in the local ISM \citep[CH$^+$ measurement in][]{sta08}. In addition to the EW uncertainties, the deviation of the $^{12}$C$^{13}$C rotational distribution from that of $^{12}$C$_{2}$, which was assumed here, can cause additional uncertainties in the resultant carbon isotope ratio. By taking higher-quality spectra and detecting multiple rotational lines (especially those arising from different rotational levels), the carbon isotope ratios of C$_{2}$ molecules can be investigated more accurately. \citet{bak98} suggested that the isotopic exchange reaction of C$_{2}$ molecules is too slow to alter the $^{12}$C$_2$ / $^{12}$C$^{13}$C ratio. The observation of $^{12}$C$_2$ and $^{12}$C$^{13}$C will give us new information about the interstellar carbon isotope ratio and related chemical processes in the regime of translucent clouds.

\section{Summary}

We reported the first detection of the C$_2$ (0,0) Phillips band and the CN (0,0) red band in the interstellar medium. We obtained the NIR high-resolution spectrum of Cyg OB2 No.\,12 using the WINERED spectrograph mounted on the 1.3 m Araki telescope in Kyoto, Japan. We could detect the (0,0) and (1,0) bands of C$_2$ and CN with high S/N. In particular, the velocity components were clearly resolved in the C$_2$ (0,0) band spectrum ($R=68,000$). Our findings are summarized as follows:

\begin{enumerate}
\item The CN column densities at levels of $N''=0$ and 1 were estimated to be $(6.7 \pm 0.3) \times$ 10$^{13}$ and $(3.3 \pm 0.3) \times$ 10$^{13}$ cm$^{-2}$, respectively. From the ratio, the rotational excitation temperature and the CN total column density were calculated as $T_{10} = 3.7 \pm 0.3$ K and $N(CN) = (8.2\pm0.4)\times 10^{13}$ $T_{10} = 3.0 \pm 0.2$ K and $N(CN) = (1.01 \pm 0.04) \times 10^{14}$ cm$^{-2}$, respectively. 

\item From the rotational distribution of C$_2$, the temperatures and densities, $(T,n)$, of components 2 and 3 were estimated to be $(30\pm5$ K, $100\pm 7$ cm$^{-3})$ and $(25\pm5$ K, $125\pm 7$ cm$^{-3})$, respectively. These essential physical parameters for understanding the interstellar chemistry could be estimated with high accuracy, due to the large oscillator strength of the (0,0) band, which allowed us to detect the rotational lines from high rotational levels for each velocity component.

\item  From the line ratios between the C$_2$ (0,0) and (1,0) Phillips bands, the oscillator strength ratio, $f_{00}/f_{10}$, of the Phillips bands were constrained. The ratio is estimated to be $f_{00}/f_{10} = 0.96 \pm 0.05$, which is consistent with both theoretical and experimental values within 1$\sigma$ uncertainties. In order to put stronger constraints on $f_{00}/f_{10}$ from the astronomical observations, it is necessary to obtain high-resolution spectra, in which the velocity components are resolved for both bands. \deleted{which favors the value theoretically derived rather than those obtained through experiments.}

\item We marginally detected the $Q(3)$ lines of $^{12}$C$^{13}$C at the velocities of components 2 and 3. If these lines are real, this is the first detection of $^{12}$C$^{13}$C in the interstellar medium. Assuming that the oscillator strength and rotational distributions of $^{12}$C$^{13}$C are the same as those of  $^{12}$C$_2$, the carbon isotope ratio was estimated to be 50--100, which is roughly consistent with the values in the local interstellar medium measured from other carbonaceous molecular features.

\end{enumerate}

In this paper, we demonstrated that the absorption bands of C$_2$ and CN in the NIR region can be used to investigate the physical properties of interstellar clouds. 
Thanks to the large oscillator strengths of these NIR bands, we could improve the accuracy of the physical parameters estimated from the rotational distributions of both C$_2$ and CN. On the other hand, the strong telluric absorption lines that overlapped with the NIR bands make the analysis complicated and increase the systematic uncertainties. The removal of telluric absorption lines is critical for analyzing these NIR bands. Along with observational improvements, the updated model of C$_2$ excitation \citep{cas12} is also important for extracting further information regarding interstellar clouds. In addition to these absorption features of small carbonaceous molecules, many DIBs, which are considered to originate from large carbonaceous molecules such as fullerenes and PAHs, have been found in the $Y$ and $J$ bands covered in our observation \citep{cox14,ham15,ham16}. In particular, the DIBs recently identified as absorption features of C$_{60}^+$ are located at about 9600\r{A}. In view of the richness of important molecular features, the wavelength range of the $Y$ and $J$ bands is crucial to the study of interstellar carbon chemistry in diffuse clouds. In the future, it will be important to investigate the relationship between DIBs (including C$_{60}^+$ features), the abundances of C$_2$ and CN and the physical parameters of interstellar clouds \citep{ely18}.

\acknowledgements

We are grateful to the staff of Koyama Astronomical Observatory for their support during our observation. 
We also thank the reviewer for a careful reading of the manuscript and helpful comments that improved this paper. 
This study is financially supported by the JSPS KAKENHI (16684001) Grant-in-Aid for Young Scientists (A), the JSPS KAKENHI (20340042) Grant-in-Aid for Scientific Research (B), the JSPS KAKENHI (26287028) Grant-in-Aid for Scientific Research (B), the JSPS KAKENHI (16K17669) Grant-in-Aid for Young Scientists (B), the JSPS KAKENHI (21840052) Grant-in-Aid for Young Scientists (Start-up), and the MEXT Supported Program for the Strategic Research Foundation at Private Universities, 2008-2012 (No. S0801061) and 2014-2018 (No. S1411028). S.H. is supported by the Grant-in-Aid for JSPS Fellows Grant No. 13J10504. N.K. is supported by JSPS-DST under the Japan-India Science Cooperative Programs during 2013-2015 and 2016-2018. K.F. is supported by the KAKENHI (16H07323) Grant-in-Aid for Research Activity start-up.

\end{document}